\documentclass[aps,twocolumn,superscriptaddress,floatfix,letterpaper,nofootinbib]{revtex4-1}

\renewcommand{\Ref}[1]{Ref.~\onlinecite{#1}}

\usepackage{amsthm}
\newtheoremstyle{wenthm}% name of the style to be used
  {3pt}% measure of space to leave above the theorem. E.g.: 3pt
  {3pt}% measure of space to leave below the theorem. E.g.: 3pt
  {\slshape}% name of font to use in the body of the theorem
  {}% measure of space to indent
  {\bfseries}% name of head font
  {:}% punctuation between head and body
  {.5em}% space after theorem head; " " = normal interword space
  {}% Manually specify head

\theoremstyle{wenthm}

\theoremstyle{definition}

\begin{document}

\begin{titlepage}

\title{Low energy effective field theories of fermion liquids and mixed
$U(1)\times \mathbb{R}^d$ anomaly }

\author{Xiao-Gang Wen} 
\affiliation{Department of Physics, Massachusetts Institute of Technology,
Cambridge, Massachusetts 02139, USA}

\begin{abstract} 

In this paper we study gapless fermionic and bosonic systems in $d$-dimensional continuum space with $U(1)$ particle-number conservation and $\mathbb{R}^d$ translation symmetry. We write down low energy effective field theories for several gapless phases where $U(1)\times \mathbb{R}^d$ is viewed as internal symmetry. The $U(1)\times \mathbb{R}^d$ symmetry, when viewed as an internal symmetry, has a mixed anomaly, and the different effective field theories for different phases must have the same mixed anomaly. Such a mixed anomaly is proportional to the particle number density, and can be measured from the distribution of the total momentum $\boldsymbol{k}_\text{tot}$ for low energy many-body states (\ie how such a distribution is shifted by $U(1)$ symmetry twist $\boldsymbol{a}$), as well as some other low energy universal properties of the systems. In particular, we write down low energy effective field theory for Fermi liquid with infinite number of fields, in the presence of both real space magnetic field and $\boldsymbol{k}$-space ``magnetic'' field. The effective field theory also captures the mixed anomaly, which constraints the low energy dynamics, such as determine the volume of Fermi surface (which is another formulation of Luttinger-Ward-Oshikawa theorem).

\end{abstract}

\maketitle

\end{titlepage}

\setcounter{tocdepth}{1} 
{\small \tableofcontents }

\section{Introduction}

A gapless Fermi liquid for spinless fermions in $d$-dimensional space is
described by the following integrated Boltzmann equation (if we ignore the
collision term that is irrelevant under the renormalization group scaling)
\cite{KLW9575}
\begin{align}
&
\prt_t u(\v x,\v k_F,t) 
+\prt_{\v x}\cdot\big(\v v_F(\v k_F) u(\v x,\v k_F,t) \big)
\nonumber\\
&
+\prt_{\v k_F} \cdot \big(\v f(\v x) u(\v x,\v k_F,t) \big)
= 0,
\end{align}
where $\v k_F$ parametrizes the Fermi surface, $\v v_F$ is the Fermi velocity,
$u(\v x,\v k_F)$ describes the Fermi surface displacement at Fermi momentum $\v
k_F$ and spacial location $\v x$, and $\v f$ is the force acting on a fermion.
We can view the integrated Boltzmann equation as the equation of motion for the
bosonized Fermi liquid.\cite{L7920,Hc0505529,HM9390,NF9493}  Together with the total
energy (assuming $\v f=0$)
\begin{align}
 E = 
\int \dd^d \v x \frac{\dd^{d-1} \v k_F}{(2\pi)^d}\, \frac{|\v v_F(\v k_F)|}{2}
u^2(\v x,\v k_F),
\end{align}
we obtain a phase-space low energy effective Lagrangian for Fermi liquid (see
\eqn{LphF1}).  Note that the low energy effective field theory contain infinite
number of fields labeled by $\v k_F$.  Or alternatively, the low energy
effective field theory can be viewed as having a single scaler field in
$(d+d-1)$-dimensional space, but the interaction in the $d-1$ dimensions
(parametrized by $\v k_F$) is allowed to be non-local.

However, such a low energy effective theory fails to capture
one of the most important properties of Fermi liquid: the volume enclosed by
the Fermi surface is $(2\pi)^d \bar \rho$ where $\bar \rho$ is the density of
the fermion in the ground state,\cite{LW6017,Oc0002392} since the fermion
density $\bar \rho$ does not even appear in the above formulation.

In recent years, it was realized that the Lieb-Schultz-Mattis (LSM)
theorem\cite{LSM6107} and its higher-dimensional generalizations by
Oshikawa\cite{Oc0002392} and Hastings\cite{H0431} can be understood in term of
a mixed anomaly between translation symmetry and an internal
symmetry.\cite{FO150307292,CB151102263,PZ170306882,LO170509298,L170504691,C180410122,JY190708596}
For a 1-dimensional system with $U(1)$ symmetry and translation symmetry, there
is a similar LSM theorem when the $U(1)$ charge per site is not an
integer.\cite{CGW1107} This suggests that such a system also has a mixed
anomaly when the $U(1)$ charge per site is not an integer.  Similarly, for
continuum systems with $U(1)$ and translation $\R^d$ symmetries, there should
also be a mixed anomaly, whenever the $U(1)$ charge density is non-zero.

The low energy effective field theory for systems with $U(1)\times \R^d$ symmetry
should capture this mixed $U(1)\times \R^d$ anomaly.  Here we like to remark
that the $U(1)\times \R^d$ symmetry is an exact symmetry.  In the low energy
effective field theory, $U(1)\times \R^d$ can be viewed as an internal symmetry.  But
when viewed as an internal symmetry, $U(1)\times \R^d$ may have a mixed
anomaly, and this is the so called mixed anomaly discussed in this paper and in
\Ref{FO150307292,CB151102263,PZ170306882,LO170509298,L170504691,C180410122,JY190708596}

The Fermi liquid at low energies also has many emergent symmetries.  In
particular, the $U(1)$ fermion-number-conservation symmetry is enlarged to
$U^\infty(1)$ emergent symmetry.\cite{L7920,Hc0505529,HM9390,NF9493}  Recently,
it was pointed out that such an emergent $U^\infty(1)$ symmetry also has an
anomaly in the presence of $U(1)$ flux.\cite{ES200707896} Using such an
$U^\infty(1)$ anomaly, one can also derive the relation between the volume
enclosed by the Fermi surface and the density of the fermion.

In this paper, we will carefully write down the low energy effective field
theories for some gapless phases of bosons and fermions.  The low energy
effective field theories contain a proper topological term that captures the
mixed $U(1)\times \R^d$ anomaly.  Such a mixed $U(1)\times \R^d$ anomaly
ensures that the system must be gapless.  

In particular, the low energy effective field theory \eq{LphF} for Fermi liquid
is obtained, that contains the proper mixed $U(1)\times \R^d$ anomaly.  Such a
mixed $U(1)\times \R^d$ anomaly determines the volume enclosed by the Fermi
surface, within the low energy effective field theory.  We also write down the
low energy effective field theory \eq{LphFBtB} for Fermi liquid with real space
magnetic field and $\v k$-space ``magnetic''
field.\cite{TKN8205,SNc9908003,XN09072021} Those are the main results of this
paper.

We will also discuss the universal low energy properties of the gapless phases
for systems with $U(1)\times \R^d$ symmetry.  Some of the features in the
universal low energy properties are determined by the mixed $U(1)\times \R^d$
anomaly, and we identify those features.

In section \ref{1dB}, we will first discuss low energy effective field theory
of 1d  weakly interacting bosons.  Then in section \ref{1dF}, we will consider
1d  weakly interacting fermions.  In section \ref{FL}, we will obtain a low
energy effective field theory for Fermi liquid in a general dimension.  Section
\ref{FpL} discusses another gapless phase of fermions -- a fermion-pair liquid,
and its low energy effective field theory.  All those effective field theories
capture the mixed $U(1)\times \R^d$ anomaly.

In this paper, we will use the natural unit where $\hbar = e = c =1$.

\section{1d boson liquid with $U(1)\times \R$ symmetry}
\label{1dB}

\subsection{Gapless phase of weakly interacting bosons}

In this section, we are going to consider 1d gapless systems in continuum space
with $U(1)$ particle-number-conservation symmetry and $\R$ translation
symmetry.  The systems are formed by bosons with weak interaction, which gives
rise to a gapless state: a ``superfluid'' state for bosons.
% or a Tomonaga-Luttinger liquid for fermions.  
We assume the system to have a size $L$ with a periodic boundary condition.  We
will compute distribution of the total momentum for many-body low energy
excitations, and how such a distribution depends on the $U(1)$ symmetry twist
described by a constant $U(1)$ background vector potential $a$.  We will see
that such a dependence directly measure a mixed anomaly in $U(1)\times \R$
symmetry, if we view $U(1)\times \R$ as an internal symmetry in the effctive
field theory.

Using the results from a careful calculation in Appendix \ref{Var}, we find the
following low energy effective field theory for the gapless phase of bosons
\begin{align}
\label{Lx1d}
  L_\text{ph} = \int \dd  x\;  &
\big(
\bar \rho \dot \phi( x,t) +
\del \rho( x,t) \dot \phi( x,t)
\\
&\ 
 - \frac {\bar \rho}{2M_b} | \prt \phi|^2 -  \frac{g}{2} \del \rho^2 +\ \cdots 
\big) ,
\nonumber\\
 L_\text{co} = \int \dd  x\;  &
\big(
\bar \rho \dot \phi( x,t) 
 - \frac {\bar \rho}{2M_b} | \prt \phi|^2 +  \frac{1}{2g} (\dot \phi)^2 +\ \cdots 
\big) ,
\nonumber 
\end{align}
where $\del \rho$ is the boson density fluctuation, $\bar \rho = \bar N/L$ the
boson density in the ground state, and $\phi$ an angular field $\phi(x,t) \sim
\phi(x,t)+2\pi$.  $L_\text{ph}$ is the phase-space Lagrangian and $L_\text{co}$
is the coordinate-space Lagrangian.  They both describe the system at low
energies.  People usually drop the total derivative term (also called
topological term) $\bar \rho \dot \phi( x,t)$, since it does not affect the
classical equation of motion of the fields.  We will see that the topological
term affects the dynamics in quantized theory and should not be dropped.

\begin{figure}[tb]
\begin{center}
\includegraphics[scale=0.4]{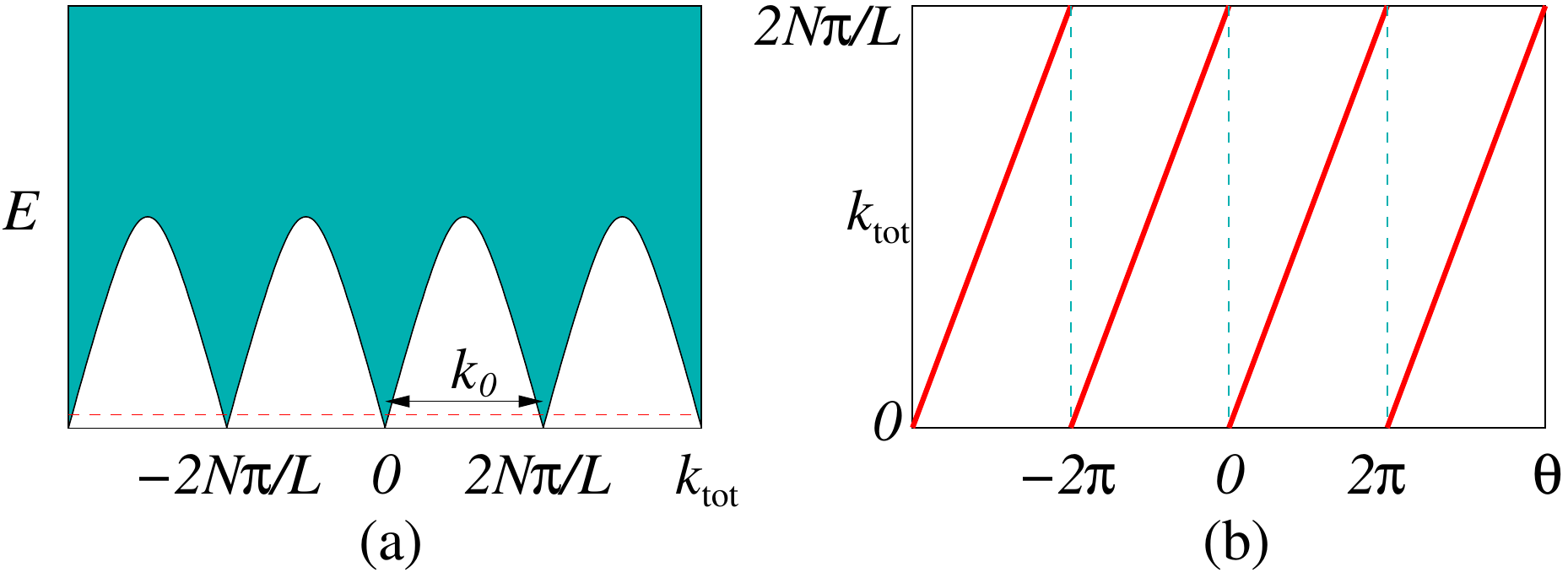} \end{center}
%Fig. 1
\caption{(a) The distribution of the total energies and total momenta for low
energy states of 1d weakly interacting boson liquid.  (b) The distribution of
the total momenta $k_\text{tot}$ for low energy many-body states, and its
dependence on the $U(1)$ symmetry twist $\th$.  } \label{aK1d}
\end{figure}

From the low energy effective field theory, we find that the low energy
excitations are labeled by $(N \in \N ,m \in \Z,n_k \in \N)$.  The total energy
and total momentum of those excitations are given by (in the presendence a
constant $U(1)$ connection $a$ describing the $U(1)$ symmetry twist)
\begin{align}
\label{EK1d}
 E &= \frac{N}{2M_b} (\frac{2\pi m}{L} + a )^2 + \frac{g}{2} (N-\bar N)^2 +
\sum_{k\neq 0} (n_{k} +\frac12) v |k|,
\nonumber\\
 k_\text{tot} &= \int\dd x\, \rho (\prt_x+a) \phi = N (\frac{2\pi m}{L} +  a ) +
\sum_{k\neq 0} n_{k} k,
\end{align}
where $N=\bar N+\del N$ is the total number of bosons in the excited state.
When $N=\bar N$ and $a=0$, the possible values of $(E,k_\text{tot})$ are
plotted in Fig.  \ref{aK1d}a.

After quantization, we have the following operator algebra
\begin{align}
\label{phirho}
 [ \phi(x), \del \rho(y) ] = \ii \del(x-y),
\end{align}
where $\del \rho(x)$ is the boson density operator.  Since
\begin{align}
 [  \del \rho(y),  \ee^{\ii \phi(x)} ] =  \del(x-y) \ee^{\ii \phi(x)},
\end{align}
$\ee^{\ii \phi(x)}$
is the boson creation operator.

Let $\vphi(x) = 2\pi \int^x \dd x' \del \rho(x')$. We find that
\begin{align}
\label{phivphi}
 [ \vphi(x), \phi(y) ] &= -2\pi \ii \Th(x-y) ,
\nonumber\\
\Th(x) &=
\begin{cases}
 1 & \ \ \text{ for } x>0 ,\\
 0 & \ \ \text{ for } x<0 ,\\
\end{cases}
\end{align}
or
\begin{align}
 \ee^{\ii \al  \vphi(x)} \ee^{\ii \bt  \phi(y)}
&= 
\ee^{\ii \bt  \phi(y)} \ee^{\ii \al  \vphi(x)} 
\ee^{ \al \bt [ \phi(y),  \vphi(x)] } 
\nonumber\\
&= 
\ee^{\ii \bt  \phi(y)} \ee^{\ii \al  \vphi(x)} 
\ee^{2 \pi \ii\al \bt \Th(x-y) } 
\end{align}
which can also be rewritten as
\begin{align}
 \ee^{-\ii \al  \vphi(x)} \ee^{-\ii \bt  \phi(y)}
 \ee^{\ii \al  \vphi(x)} \ee^{\ii \bt  \phi(y)}
&=\ee^{2 \pi \ii\al \bt \Th(x-y) } ,
\nonumber\\
\ee^{-\ii \bt  \phi(y)}
 \ee^{-\ii \al  \vphi(x)} 
\ee^{\ii \bt  \phi(y)}
 \ee^{\ii \al  \vphi(x)} 
&=\ee^{-2 \pi \ii\al \bt \Th(x-y) } .
\end{align}

The above expression tells us that that the operator $\ee^{\ii \al \vphi(x)}$
cause a $\ee^{-2\pi \ii\al}$ phase shift for the operator $\ee^{\ii  \phi(y)}$
for $y< x$, and keep $\ee^{\ii  \phi(y)}$ unchanged for $y>x$.  So the operator
$\ee^{\ii  \vphi(x)}$ increases $m$ by 1, \ie increase the total momentum by
$2\pi \bar \rho$.  Similarly, the operator $\ee^{\ii \bt \phi(x)}$ cause a
$\ee^{2\pi \ii\bt}$ phase shift for the operator $\ee^{\ii  \vphi(y)}$ for $y<
x$, and keep $\ee^{\ii  \vphi(y)}$ unchanged for $y>x$.  So the operator
$\ee^{\ii  \phi(x)}$ increases $N=\frac{1}{2\pi} \int \dd x\, \prt_x \vphi =
\int \dd x\, \del \rho$ by 1.

From the above results, we see that under the $U(1)$ transformation $\th$
\begin{align}
 \phi(x) \to \phi(x)+\th,\ \ \ \ \vphi(x) \to \vphi(x) .
\end{align}
Under the $\R$ translation transformation $\del x$
\begin{align}
 \phi(x) \to \phi(x+\del x),\ \ \ \ \vphi(x) \to \vphi(x+\del x) + 2\pi \bar \rho \del x .
\end{align}
However, at low energies, for smooth fields and small $\del x$, we have
$\phi(x) \approx \phi(x+\del x)$ and $\vphi(x) \approx \vphi(x+\del x)$.  Thus
under the $\R$ transformation $\del x$
\begin{align}
 \phi(x) \to \phi(x),\ \ \ \ \vphi(x) \to \vphi(x) + 2\pi \bar \rho \del x .
\end{align}
This way, the $\R$ symmetry becomes an internal symmetry in the low energy
effective field theory \eq{Lx1d}.  We will see that $U(1)\times \R$, when
viewed as an internal symmetry, has a mixed anomaly.

We note that $\ee^{\ii \vphi(x)}$ is a local operator and $\int \dd x\, \del
\rho = \frac{1}{2\pi} \int \dd x\, \prt_x \vphi$ is an integer. Both imply that
$\vphi$ is also an angular variable $\vphi(x,t) \sim \vphi(x,t) + 2\pi$.  Using
the two angular fields $\phi_1=:\phi$ and $\phi_2=:\vphi$, the low energy effective theory can
be written as
\begin{align}
\label{LK}
  L_\text{ph} = \int \dd  x\;  \big(
\bar \rho \prt_t \phi_1 + \frac{K_{IJ}}{4\pi} \prt_x \phi_I \prt_t \phi_J
- \frac{V_{IJ}}{2} \prt_x \phi_I  \prt_x \phi_J \big)
\end{align}
where 
\begin{align}
K = \begin{pmatrix}
 0&1\\
 1&0\\
\end{pmatrix} .
\end{align}
and $V$ is a positive definite symmetric matrix.

\subsection{The universal properties of the gapless phase}

From \eqn{EK1d}, we see that the total momentum
$k_\text{tot}$ depends on the $U(1)$ symmetry twist
\begin{align}
k_\text{tot} = \rho \th,\ \ \ \text{ where } 
 \th = \int \dd x\; a = aL,\ \ \rho = \frac{N}{L}.
\end{align}
However, since the state is gapless, there are many low energy states with
different momenta. So we do not know $k_\text{tot}$ to be the momentum of which
low energy states.  To make our statement meaningful, we consider two low
energy states $|\Psi_{N}\>$ of $N$ bosons.  Let $k_\text{tot}$ be the total
momenta of $|\Psi_{N}\>$.  Since there are many different low energy states
$|\Psi_{N}\>$'s, we have many different values of $k_\text{tot}$.  In other
words, we have a distribution of $k_\text{tot}$'s.  Such a distribution is
plotted in Fig.  \ref{aK1d}b.

From the distribution pattern in Fig. \ref{aK1d}, we see two universal
properties: the period in the distribution and the $\th=aL$ dependence of the
distribution
\begin{align}
\label{UP1d}
 k_0 = 2\pi \bar \rho,\ \ \ \ \ \frac{\dd k_\text{tot}}{\dd \th} = \bar \rho, 
\end{align}
which do not depend on the small changes in the interactions and the dispersion
of the bosons, unless those changes cause a phase transition.  Thus we say they
are universal properties that characterize the gapless phase.  The two
universal properties are closely related $(2\pi)^{-1}k_0 = \frac{\dd
k_\text{tot}}{\dd \th} = \bar\rho$.  We call $\bar\rho$ an index for the
gapless phase.  Physically $\bar\rho$ is nothing but the density of the $U(1)$
charges in the ground state.

Let us give a argument why $ \frac{\dd k_\text{tot}}{\dd \th} $ is
universal.  Let us assume the $U(1)$ symmetry twist is described by a boundary
condition on single-particle wave function at $x_0$: $\psi(x_0+0^+) = \ee^{\ii
\th} \psi(x_0-0^+)$.  A usual translation $x \to x+\Del x$ will shift the
symmetry twist from $x_0$ to $x_0+\Del x$.  So the symmetry twist breaks the
translation symmetry.  But we can redefine the translation operator to be the
usual translation plus a $U(1)$ transformation $\psi(x) \to \ee^{\ii \th}
\psi(x)$ for $x \in [x_0,x_0+\Del x]$.  The new translation operator generates
the translation symmetry in the presence of the $U(1)$ symmetry twist.  Due to
the $U(1)$ transformation $\psi(x) \to \ee^{\ii \th} \psi(x)$ for $x \in
[x_0,x_0+\Del x]$, the eigenvalue of the new translation operator has a $\th$
dependence given by $(\ee^{\ii \th} )^{\bar\rho \Del x}$, where $\bar\rho \Del
x$ is the total $U(1)$ charges in the interval $[x_0,x_0+\Del x]$.  In other
words the total momentum has a $\th$ dependence given by $\th \bar\rho$.  This
is the reason why $ \frac{\dd k_\text{tot}}{\dd \th} =\bar\rho $.  Since
$\th=0$ and $\th=2\pi$ are equivalent, therefore $ \frac{\dd
k_\text{tot}}{\dd \th} =\bar\rho $ implies the periodicy in Fig. \ref{aK1d},
with the period $k_0 =2\pi \bar\rho$.  The above discussion does not depend on
interactions and boson dispersion. So the results \eq{UP1d} are universal
properties.

\subsection{Mixed anomaly for $U(1)\times \R$ symmetry}

From the above argument, we also see that the shift of the low energy momentum
distribution by the $U(1)$ symmetry twist, $\frac{\dd k_\text{tot}}{\dd
\th} =\bar\rho $, is an invariant not only against small perturbations that
preserve the $U(1)\times \R$ symmetry, but is also an invariant against large
symmetry-preserving perturbations that can drive through a phase transition.
The invariant for large perturbations is actually an anomaly.\cite{H8035} This
is because, under a new point of view,\cite{W1313,KW1458} an anomaly
corresponds to an SPT or topological order in one higher
dimension.\cite{W1313,KW1458}  Any large perturbations and phase transitions
cannot change the SPT or topological order in one higher dimension.

In our case, we can view $\frac{\dd k_\text{tot}}{\dd \th} =\bar\rho $ as
an anomaly in the low energy effective field theories \eq{Lx1d} and \eq{LK}.
We see that the topological term $\bar \rho( x,t) \dot \phi( x,t)$ in the low
energy effective field theory determine the anomaly.  In fact, such an anomaly
is a mixed anomaly between $U(1)$ symmetry and the translation $\R$ symmetry,
which describe how an $U(1)$ symmetry twist can change the total momentum (\ie
$\frac{\dd k_\text{tot}}{\dd \th}\neq 0$).

The presence of the anomaly implies that the ground state of the system must be
either gapless or have a non-trivial topological order.  Since there is no
non-trivial topological order in 1d, the  ground state must be gapless.  In
other words, \frmbox{the field theories in \eqn{Lx1d} and \eqn{LK} with $\bar
\rho \neq 0$ must be gapless regardless the interaction term described by
$\cdots$, as long as the $U(1)\times \R$ symmetry is preserved.  On the other
hand, when $\bar \rho = 0$, the field theories in \eqn{Lx1d} and \eqn{LK} allow
a gapped phase with  $U(1)\times \R$ symmetry.  }

The mixed anomaly between the  $U(1)$ symmetry and the $\R$ symmetry can also
be detected via the patch symmetry transformations studied in
\Ref{JW191213492}.  The $U(1)$ patch symmetry transformations are given by
\begin{align}
\label{WU1}
 W_{U(1)}(x,y)
= \ee^{\ii 2\pi \al  \int_x^y \dd x \rho}
= \ee^{\ii \al(\vphi(y) -\vphi(x)) }
 \ee^{\ii 2\pi \al  \bar \rho (y-x) },
\end{align}
which perform the $U(1)$ transformation, $\phi \to \phi+\al$, on the segment
$[x,y]$.  The $\R$ patch symmetry transformations are given by
\begin{align} 
\label{WR}
 W_{\R}(x,y) 
= \ee^{  \ii \bar \rho \Del x (\phi(y) -\phi(x)) }.
\end{align}
which perform the $\R$ transformation (the translation $\Del x$) on the segment
$[x,y]$.  In the low energy limit, $k \to 0$. So for a finite $\Del x$ the
translation is trivial for the phonon modes.  The translation $\Del x$ has a
non-trivial actions only on sector labeled by different $N$'s and $m$'s.  For a
translation $\Del x$ that acts on a segment $[x,y]$, its effect is to transfer
$U(1)$ charge-$\bar \rho \Del x$ from $x$ to $y$.  This is why the $\R$ patch
symmetry transformations are given by \eqn{WR}.

In the low energy effective theories \eq{Lx1d} and \eq{LK}, the $U(1)$
transformation is given by $\phi \to \phi+\th$.  The term $\bar \rho \dot \phi$
implies $\bar \rho$ is the background $U(1)$ charge density.  Therefore, the
patch translation transformation has a form \eqn{WR}.

Assume $x_2>x_1$. We have shown that $W_{U(1)}(x_1,x_2)$ shifts $\ee^{  \ii \bt
\phi(y)}$ by a phase $\ee^{ -\ii 2\pi \al \bt }$, if $x_1 < y < x_2$.
Therefore
\begin{align}
\label{WU1WR}
&\ \ \ \
 W_{U(1)}(x_1,x_2)  W_{\R}(y_1,y_2) 
\nonumber\\
&= W_{\R}(y_1,y_2) W_{U(1)}(x_1,x_2)  
\ee^{\ii 2\pi \al  \bar \rho \Del x } ,
\end{align}
for $x_1 < y_1 < x_2 < y_2$.  The extra phase factor $\ee^{\ii 2\pi \al  \bar
\rho \Del x} $ indicates the appearance of the mixed $U(1)\times \R$ anomaly.
Using the terminology of \Ref{JW191213492}, we say, the $U(1)$ symmetry and the
$\R$ symmetry have a ``mutual statistics'' between them, as a consequence of the
mixed anomaly. So, according to \Ref{JW191213492}, the $U(1)$ symmetry and the
$\R$ are not independent, and we may denote the combined
symmetry as $U(1)\vee \R$ to stress the mixed anomaly.

We like to remark that $\phi$ and $\vphi$ fields are cannonial conjugate to
each other.  We see that the symmetries that shift $\phi$ and $\vphi$ have a
mixed anomaly, as captured by non-trivial commutation relation between the
patch operators for the symmetry transformation.  This is a general mechanism
of the appearance of anomaly.

\section{1d fermion liquid with $U(1)\times \R$ symmetry}
\label{1dF}

\subsection{Weakly interacting 1d gapless fermionic systems}

1d gapless fermionic systems with $U(1)\times \R$ symmetry and weak repulsive
interaction are also in a gapless phase -- a Tomonaga-Luttinger liquid for
fermions.  The low energy effective theory also has a form
\begin{align}
\label{Lx1dF}
  L_\text{ph} = \int \dd  x\;  &
\big(
\bar \rho \dot \phi( x,t) +
\del \rho( x,t) \dot \phi( x,t)
\nonumber\\
&
 - \frac {\bar \rho}{2M_f} | \prt \phi|^2 +  \frac{1}{2g} (\dot \phi)^2 +\ \cdots 
\big) .
\end{align}
Considering non-interacting fermions in a system with periodic boundary
condition on a ring of size $L$, we find that the low energy excitations are
also labeled by $(N \in \N ,m \in \Z,n_k \in \N)$.  However, the total energies
and total momenta of those excitations are given by, for $N=$ odd,
\begin{align}
\label{EK1dO}
& E = \frac{N}{2M_f} (2\pi \frac{m}{L} + a )^2 + \frac{g}{2} \del N^2 +
\sum_{k\neq 0} (n_{k} +\frac12) v |k|,
\nonumber\\
& k_\text{tot} = N (2\pi \frac{m}{L} +  a ) +
\sum_{k\neq 0} n_{k} k,
\end{align}
and for $N=$ even,
\begin{align}
\label{EK1dE}
& E = \frac{N}{2M_f} (2\pi \frac{m+\frac12}{L} + a )^2 + \frac{g}{2} \del N^2 +
\sum_{k\neq 0} (n_{k} +\frac12) v |k|,
\nonumber\\
& k_\text{tot} = N (2\pi \frac{m+\frac12}{L} +  a ) +
\sum_{k\neq 0} n_{k} k,
\end{align}
where $N (2\pi \frac{m}{L} +  a )$ or $N (2\pi \frac{m+\frac12}{L} +  a )$ are
the momentum of center of mass.

\begin{figure}[tb]
\begin{center}
\includegraphics[scale=0.4]{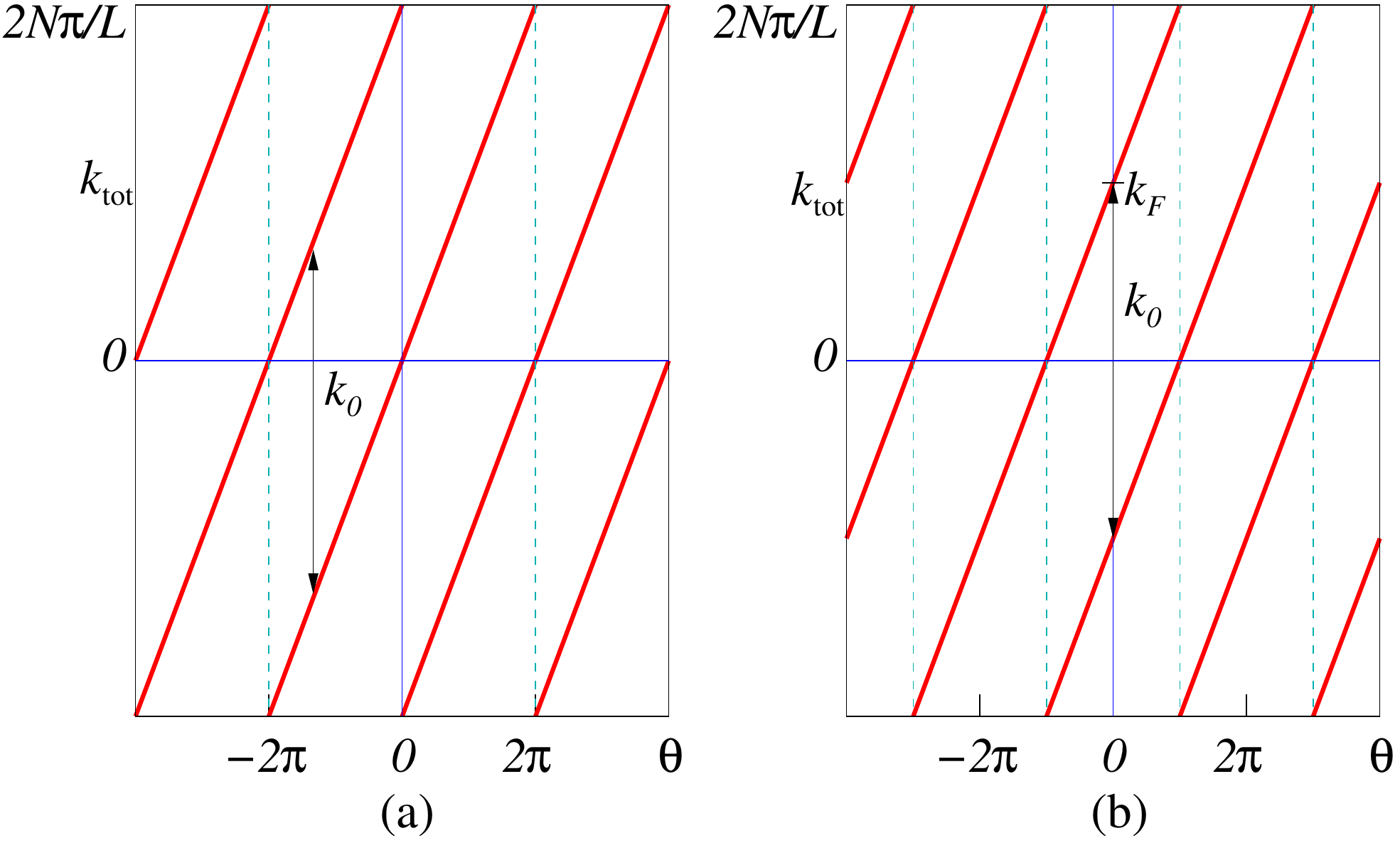} \end{center}
%Fig. 1
\caption{The distribution of the total momenta $\Del k_\text{tot}$ and its
dependence on the $U(1)$ symmetry twist, for a weakly interacting 1d fermion
liquid.  } \label{aK1dBF}
\end{figure}

Again, we consider low energy states $|\Psi_{N}\>$ of $N$ fermions.  Let
$k_\text{tot}$ be the total momenta of $|\Psi_{N}\>$.  Since there are many
different low energy states $|\Psi_{N}\>$'s,  we have a distribution of
$k_\text{tot}$'s.  Such a distribution is plotted in Fig.  \ref{aK1dBF}a for
$N_ = $ odd case, and in Fig. \ref{aK1dBF}b for $N=$ even case.  We see that
$N=$ odd case and $N=$ even case have different distributions for
$k_\text{tot}$'s.

The shift of the distribution of $k_\text{tot}$ by the $U(1)$ symmetry twist
(see Fig. \ref{aK1dBF}) can be interpreted as $U(1)$ symmetry twist producing
or pumping momentum.  This directly measures the mixed $U(1)\times \R$ anomaly.

We also see that a local operator that create a fermion (\ie change $N$ by 1)
must also change $m$ by $\frac12$.  Those operators have a form $\ee^{\ii (\phi
\pm \frac12\vphi)}$, where $\prt_x \vphi =2\pi \del\rho$.  Since the allowed
operators are generated by $\ee^{\ii (\phi \pm \frac12 \vphi)}$, the following
fields
\begin{align}
 \phi_1 = \phi + \frac12 \vphi,\ \ \ \ \ \phi_2 = \phi - \frac12 \vphi
\end{align}
are angular fields: $\phi_i \sim \phi_i +2\pi$.
Using
\begin{align}
 \phi = \frac12(\phi_1+\phi_2),\ \ \ \ \ \vphi = \phi_1-\phi_2,
\end{align}
we find the low energy effective theory to be
\begin{align}
\label{Lx1dphi}
  L &= 
\int \dd  x\; \big( 
\frac {\bar \rho}{2} \prt_t (\phi_1 +\phi_2)
\nonumber\\
&\ \ \ \ \ \ \ \ + \frac{1}{4\pi} \prt_x (\phi_1-\phi_2) \prt_t (\phi_1+\phi_2)
 - \frac {V_{IJ}}{2}  \prt_x \phi_I \prt_x \phi_J  \big)
\nonumber\\
&= \int \dd  x\; \big( 
\frac {\bar \rho}{2} \prt_t (\phi_1 +\phi_2)
+ \frac{1}{4\pi} (
\prt_x \phi_1 \prt_t \phi_2
-\prt_x \phi_2 \prt_t \phi_1
)
\nonumber\\
&\ \ \ \  \ \ \ \ + \frac{K_{IJ}}{4\pi} \prt_x \phi_I \prt_t \phi_J
 - \frac {V_{IJ}}{2}  \prt_x \phi_I \prt_x \phi_J  \big)  ,
\nonumber\\
K &= \begin{pmatrix}
1&0\\
0&-1\\
\end{pmatrix} ,\ \ \ \ 
V = \text{positive definite}.
\end{align}
Here, we have been careful to keep the total derivative terms $\frac {\bar
\rho}{2} \prt_t (\phi_1 +\phi_2) + \frac{1}{4\pi} ( \prt_x \phi_1 \prt_t \phi_2
-\prt_x \phi_2 \prt_t \phi_1) $. Those are topological terms that do not affect
the classical equation of motion, but have effects in quantum theory.

Effective theory similar to the above form has been obtained before for edge
state of fractional quantum Hall states.\cite{W9211,W9505} But here we have to
be more careful in keeping the topological term $\bar \rho \prt_t \phi_1$,
which describes the mixed anomaly of $U(1)\times \R$ symmetry for the fermionic
system.

We like to mention that the patch symmetry transformations are determined from
the low energy effective theories \eq{Lx1dF} or \eq{Lx1dphi}, and are still
given by \eqn{WU1} and \eqn{WR}.  So the mixed anomaly can still be detected via
commutation relation of the patch symmetry transformations \eq{WU1WR}.

In fact, \eqn{Lx1dphi} is the low energy effective theory for a fermion system
with Fermi momentum $k_F = \pi \bar \rho$.  $\frac1{2\pi} \prt_x \phi_1$
describes the density of right-moving fermions and $\frac1{2\pi} \prt_x \phi_2$
describes the density of left-moving fermions.
For example, the low energy effective theory for
right-moving fermions is given by
\begin{align}
\label{LRF}
&  L 
= \int \dd  x\; \big(  \frac {k_F}{2\pi} \prt_t \phi_1 
 + \frac{1}{4\pi} \prt_x \phi_1 \prt_t \phi_1
 - \frac {V_{11}}{2}  \prt_x \phi_1 \prt_x \phi_1 \big)  ,
\nonumber\\
&= \int \dd  x\; \big(  \bar\rho_1 \prt_t \phi_1 
 + \frac{1}{4\pi} \prt_x \phi_1 \prt_t \phi_1
 - \frac {V_{11}}{2}  \prt_x \phi_1 \prt_x \phi_1 \big)  .
\end{align}
In the above expression, we stress the direction connection between the
topological term and the Fermi momentum $k_F$, as well as the direction
connection between the topological term and the density of the right-moving
fermions: $\bar \rho_1 = \frac {k_F}{2\pi} = \frac12 \bar \rho$.

We see that the mixed anomaly of $U(1)\times \R$ symmetry is nothing but a
non-zero Fermi momentum $k_F$.  When $k_F=0$, \ie when  mixed anomaly vanishes,
the  fermion system can have a gapped ground state that does not break the
$U(1)\times \R$ symmetry.  But when $k_F\neq 0$, \ie in the presence of  mixed
anomaly, the  fermion system cannot have a gapped ground state that does not
break the $U(1)\times \R$ symmetry.  This is a well known result, but restated
in terms of mixed anomaly of $U(1)\times \R$ symmetry.

\section{Low energy effective theory of $d$-dimensional Fermi liquid 
and the mixed $U(1)\times \R^d$ anomaly}
\label{FL}

\subsection{Effective theory for the Fermi surface dynamics} 
\label{effFermi}

In the last section, we discussed the low energy effective theory of 1d Fermi
liquid, which contain a proper topological term that reflects the mixed
$U(1)\times \R$ anomaly.  In this section, we are going to generalize this
result to higher dimensions.  The generalization is possible since the higher
dimensional  Fermi liquid can be viewed as a collection of  1d Fermi liquids.

Let us use $\v k_F$ to parametrize the Fermi surface.  We introduce $u(\v
x,\v k_F)$ to describe the shift of the Fermi surface.  Thus the total fermion
number is given by
\begin{align}
 N = \bar N + \int \dd^d \v x \frac{\dd^{d-1} \v k_F}{(2\pi)^d}\, 
u(\v x,\v k_F).
\end{align}
The total energy is
\begin{align}
\label{Eu}
 E = \bar E + \int \dd^d \v x \frac{\dd^{d-1} \v k_F}{(2\pi)^d}\, \frac{|\v v_F(\v k_F)|}{2}
u^2(\v x,\v k_F),
\end{align}
where
\begin{align}
 \v v_F =: \prt_{\v k}H(\v k),
\end{align}
is the Fermi velocity and $H(\v k)$ is the single fermion energy.

The equation of motion for the field $u(\v x,\v k_F)$ is given by
\begin{align}
\label{EOMu}
(\prt_t +  \v v_F \cdot \prt_{\v x}) u(\v x,\v k_F) = 0
\end{align}
Let us introduce a field $\phi(\v x,\v k_F)$ via
\begin{align}
- \v n_F \cdot \prt_{\v x} \phi(\v x,\v k_F) = u(\v x,\v k_F),
\end{align}
where
\begin{align}
 \v n_F =: \frac{\v v_F}{|\v v_F|} .
\end{align}
The equation of motion for $\phi$ becomes
\begin{align}
\label{EOMf}
(\prt_t +  \v v_F \cdot \prt_{\v x}) 
\big( \v n_F \cdot \prt_{\v x} \phi(\v x,\v k_F) \big)
= 0
\end{align}
and the total energy becomes
\begin{align}
\label{Ef}
 E = \bar E + \int \dd^d \v x \frac{\dd^{d-1} \v k_F}{(2\pi)^d}\, 
\frac{|\v v_F(\v k_F)|}{2}
\v n_F \cdot \prt_{\v x} \phi (\v x,\v k_F)]^2.
\end{align}
The phase-space Lagrangian that produces the above equation of motion and total
energy is given by
\begin{align}
\label{LphF1}
L_\text{ph} = 
 \int \dd^d \v x & \frac{\dd^{d-1} \v k_F}{(2\pi)^d}\, 
\Big( -\frac12  
\prt_t \phi (\v x,\v k_F)
\v n_F \cdot \prt_{\v x} \phi (\v x,\v k_F)
\nonumber\\
& -
\frac{|\v v_F(\v k_F)|}{2}
[\v n_F \cdot \prt_{\v x} \phi (\v x,\v k_F)]^2
\Big)
\end{align}

Repeating a calculation similar to 1d chiral Luttinger
liquid,\cite{W9211,W9505} we find that after quantization, the operator $ \phi
(\v x,\v k_F)$ has the following commutation relation
\begin{align}
&\ \ \ \
 [\phi(\v x',\v k_F'), \v n_F \cdot \prt_{\v x} \phi (\v x,\v k_F)]
\nonumber\\
& =
- [\phi(\v x',\v k_F'),  u (\v x,\v k_F)]
\nonumber\\
&= -\ii (2\pi)^d 
\del^d(\v x-\v x')
\del^{d-1}(\v k_F-\v k_F')
\end{align}
which reproduce the equation of motion
\begin{align}
& \prt_t \phi(\v x,\v k_F,t)
= \ii [ H,  \phi(\v x,\v k_F,t)] = - \v v_F \cdot \prt_{\v x} \phi (\v x,\v k_F,t),
\nonumber\\
& H  =
\int \dd^d \v x  \frac{\dd^{d-1} \v k_F}{(2\pi)^d}\, 
\frac{|\v v_F(\v k_F)|}{2}
[\v n_F \cdot \prt_{\v x} \phi (\v x,\v k_F)]^2
\end{align}
We also see that
\begin{align}
 [ N,  \phi(\v x,\v k_F)] = -\ii ,\ \ \ 
 [ N,  \ee^{\ii \phi(\v x,\v k_F)}] = \ee^{\ii \phi(\v x,\v k_F)}.
\end{align}
Thus $\ee^{\ii \phi(\v x,\v k_F,t)}$ is the operator that increases $N$ by 1,
and the $U(1)$ symmetry transformation is given by
\begin{align}
\label{U1trans}
\ee^{\ii \th N}  \phi(\v x,\v k_F) \ee^{-\ii \th N} = 
\phi(\v x,\v k_F) +\th.
\end{align}
We see that $\phi(\v x,\v k_F)$ is an angular field $\phi(\v x,\v k_F) \sim
\phi(\v x,\v k_F) +2\pi$.

However, in \eqn{LphF1} we only have terms that $\phi(\v x,\v k_F) $ couples to
$\del N$.  In a compete Lagrangian, $\phi(\v x,\v k_F) $ must also couple to
$\bar \rho$ -- the density of the $U(1)$ charge in the ground state.  The
complete phase-space Lagrangian is given by
\begin{align}
\label{LphF}
L_\text{ph}  = 
 \int  \dd^d \v x & \frac{\dd^{d-1} \v k_F}{(2\pi)^d}\, 
\Big(  
\frac{\bar\rho}{A_F}\dot \phi (\v x,\v k_F)
\nonumber\\
& -\frac12  \dot \phi (\v x,-\v k_F) \v n_F \cdot \prt_{\v x} \phi (\v x,\v k_F)
\nonumber\\
& -\frac12  \dot \phi (\v x,\v k_F) \v n_F \cdot \prt_{\v x} \phi (\v x,\v k_F)
\nonumber\\
& - \frac{|\v v_F(\v k_F)|}{2}
[\v n_F \cdot \prt_{\v x} \phi (\v x,\v k_F)]^2
\Big)
\end{align}
where
\begin{align}
 A_F =:  \int \frac{\dd^{d-1} \v k_F}{(2\pi)^d} .
\end{align}
and we have assumed a central reflection symmetry $\v k_F \to -\v k_F$, \ie $\v
v_F(\v k_F) = -\v v_F(-\v k_F)$ and $\v n_F(\v k_F) = -\v n_F(-\v k_F)$.  The
two terms, $ \frac{\bar\rho}{A_F}\dot \phi (\v x,\v k_F)$ and 
\begin{align}
\int \frac{\dd^{d-1} \v k_F}{(2\pi)^d}\, \frac12
\dot \phi (\v x,-\v k_F) \v n_F(\v k_F) \cdot \prt_{\v x} \phi (\v x,\v k_F)
, 
\end{align}
are total derivative topological terms.

We know that the volume enclosed by the Fermi surface is directly related to
the fermion density $\bar \rho$.\cite{LW6017,Oc0002392}  Naively, in our
effective theory \eq{LphF}, the parameter $\bar \rho$ and Fermi surface $\v
k_F$ are not related.  In the following, we like to show that in fact $\bar
\rho$ and the Fermi surface $\v k_F$ are related, from within the effective
field theory \eq{LphF}.  

Consider a field configuration
\begin{align}
\label{phidk}
  \phi (\v x,\v k_F) & = \v a \cdot \v x
\nonumber\\
\text{or } \ \ 
 \v n_F \cdot \prt_{\v x}  \phi (\v x,\v k_F) &= \v a \cdot \v n_F = u(\v x,\v k_F).
\end{align}
There are two ways to compute the
momentum for such a field configuration.

In the first way, the total momentum is computed via the deformation $u(\v x,\v
k_F)$ of the Fermi surface (assuming the  total momentum
of the ground state to be zero)
\begin{align}
\v k_\text{tot} &= 
\int \dd^d \v x \frac{\dd^{d-1} \v k_F}{(2\pi)^d}\, \v k_F u(\v x,\v k_F)
\nonumber\\
&= \int \dd^d \v x \frac{\dd^{d-1} \v k_F}{(2\pi)^d}\, 
\v k_F (\v a \cdot \v n_F) 
.
\end{align}
Note that $\v n_F$ is the normal direction of the Fermi surface.
Therefore
\begin{align}
& \v k_\text{tot} = 
 \int \dd^d \v x \frac{\dd^{d-1} \v k_F}{(2\pi)^d}\, 
\v k_F (\v a \cdot \v n_F) 
\\
&=
\int \dd^d \v x \int_{\v k \in \v k_F+\v a} \frac{\dd^d \v k_F}{(2\pi)^d}\, \v k 
-\int \dd^d \v x \int_{\v k \in \v k_F} \frac{\dd^d \v k_F}{(2\pi)^d}\, \v k 
,
\nonumber 
\end{align}
where $\int_{\v k \in \v k_F} \dd^d \v k_F$ means integration over $\v k$
inside the Fermi surface, and $\int_{\v k \in \v k_F+\v a} \dd^d \v k_F$
means integration over $\v k$ inside the shifted Fermi surface (shifted by
$\v a$).  Let 
\begin{align}
 V_{\v k_F} =: \int_{\v k \in \v k_F} \dd^d \v k_F
\end{align}
be the volume enclosed by the Fermi surface,
we see that
\begin{align}
\label{kVF}
 \v k_\text{tot} = V \frac{V_{\v k_F}}{(2\pi)^d} \v a 
%= V \bar \rho \v a
%= \bar N \v a,
\end{align}
where $V$ is the volume of the system.
% and $\frac{V_{\v k_F}}{(2\pi)^d} =\bar \rho$ is the density of the fermions.

There is a second way to compute the total momentum $\v k_\text{tot}$.  We
consider a time dependent translation of the above configuration
\begin{align}
  \phi (\v x,\v k_F) & = \v a \cdot (\v x +\v x_0(t)).
\end{align}
The effective phase-space Lagrangian for $\v x_0(t)$ is given by
\begin{align}
 L_\text{ph} = V \bar \rho \v a \cdot \dot {\v x}_0(t) .
\end{align}
We see that $V \bar \rho \v a$ is the canonical momentum of the
translation $\v x_0$.  Thus, the total momentum of the configuration is
\begin{align}
\label{krho}
 \v k_{tot} =  V \bar \rho \v a.
\end{align}
Compare \eqn{kVF} and \eqn{krho}, we see that the volume included by the Fermi
surface and the fermion density is related
\begin{align}
\label{VkFrho}
\frac{V_{\v k_F}}{(2\pi)^d}  = \bar\rho .
\end{align}
This is the Luttinger theorem.

\subsection{The mixed anomaly in $U(1)\times \R^d$ symmetry}

As we have pointed out that the topological term $\int \dd^d \v x
\frac{\dd^{d-1} \v k_F}{(2\pi)^d}\, \big( \frac{\bar\rho}{A_F}\prt_t \phi (\v
x,\v k_F) $ represents a mixed anomaly of $U(1)\times \R^d$ symmetry.  To see
this point, we note that $\v a$ in \eqn{phidk} can be viewed as the $U(1)$
symmetry twist.  The fact that the $U(1)$ symmetry twist can induce the $\R^d$
quantum number (\ie the momentum) reflects the presence of the mixed anomaly of
$U(1)\times \R^d$ symmetry.  Eqn. \eq{VkFrho} indicates that the mixed anomaly
can constraint the low energy dynamics, in this case determines the volume
enclosed by the Fermi surface.

\begin{figure}[tb]
\begin{center}
\includegraphics[scale=0.25]{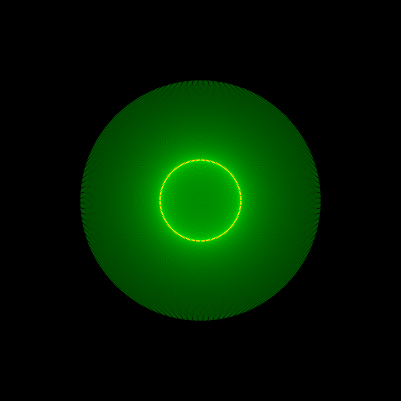} ~~~~
\includegraphics[scale=0.25]{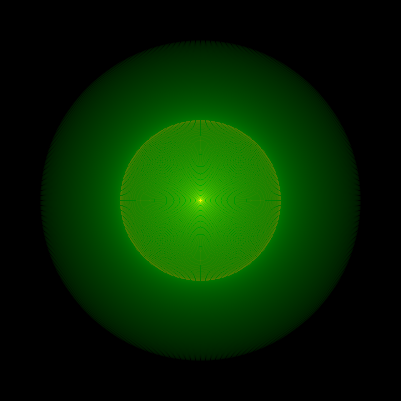} 
\end{center}
%Fig. 1
\caption{
The total momentum distributions for low energy
many-body states for a 2d Fermi liquid, 
with 1-particle excitations (left, red channal), 
2-particle 1-hole excitations (left, green channal),
1-particle 1-hole excitations (right, red channal), 
and 2-particle 2-hole excitations (right, green channal).
The horizental axis is $k_x$ and the vertical axis is $k_y$.  
} \label{1kF3kF}
\end{figure}

In the above, we discussed how the $U(1)$ symmetry twist shifts the total
momentum of a particular low energy many-body state.  However, in practice, we
cannot pick a particular low energy many-body state, and see how its momentum
is shifted by the $U(1)$ symmetry twist.  What can be done is to examine all
the low energy many-body low energy states, and their total momentum
distribution.  The shift of the total momentum distribution by the $U(1)$
symmetry twist measure the mixed $U(1)\times \R^d$ anomaly.  In Fig.
\ref{1kF3kF}, we plot the total momentum distributions for low energy
many-body states, with 1-particle excitations, 2-particle 1-hole excitations,
1-particle 1-hole excitations, and 2-particle 2-hole excitations.

The  mixed anomaly not only appears in Fermi liquid phases of fermions, it also
appears in any other phases of fermions. Thus the  mixed anomaly constrain the
low energy dynamics in any of those phases.  In next section, we consider a
phase of fermion, where fermions pair-up to form a boson liquid in
$d$-dimensional space.

\subsection{Effective theory of a Fermi liquid in most general setting}

In section \ref{effFermi}, we considered Fermi liquid in free space.  In this
section, we like to include electromagnetic field in real space, as well as
``magnetic field'' in $\v k$ space.  We like to find the low energy effective
theory of Fermi liquid for this more general situation.

First we consider the dynamics of a single particle in a very general setting.
The classical state of the particle is described by a point in phase-space
parametrized by $\xi^I$.  The single particle dynamics is described by a
single-particle phase-space Lagrangian:
\begin{align}
\label{phLxi}
 L(\dot \xi^I,\xi^I) 
= \int \dd t\ \big[ a_I(\xi^I) \dot \xi^I - H(\xi^I)\big],
\end{align}
which gives rise to the following single-particle equation of motion
\begin{align}
\label{xiHb}
 b_{IJ}\dot \xi^J &= \frac{\prt  H}{\prt \xi^I}, \ \ \ \ \ \ 
b_{IJ}=\prt_{\xi^I} a_J -\prt_{\xi^J} a_I .
\end{align}
Here $H(\xi^I)$ is the single-particle energy for the state $\xi^I$, and
$a_I(\xi^I)$ is a phase-space vector potential that describes the phase
``magnetic'' field $b_{IJ}(\xi^I)$.  The phase space ``magnetic'' field
includes both the real space magnetic field and $\v k$-space  ``magnetic''
field.\cite{TKN8205,SNc9908003,XN09072021}  

For a particle in a $d$-dimensional free space described by coordinate-momentum
pair $(\v x, \v k) = (x^i,k_i)$, $i=1,\cdots,d$, the phase-space magnetic field
$b_{IJ}(\xi^I)$ is a constant (\ie independent of $\xi^I$), since $ a_I(\xi^I)
\dot \xi^I = \v k \cdot \dot{\v x}$ (\ie $a_{x^i} = k_i,\ a_{k_i}=0$).  On the
other hand, if there is a non-uniform real space and/or $\v k$-space magnetic
fields, phase-space ``magnetic'' field $b_{IJ}(\xi^I)$ will not be uniform.

As a example, let us consider a particle in 3-dimensional space.  The phase
space is 6-dimensional and is parametrized by $(\xi^I)=(\v x,\v k)$.  The
phase-space Lagrangian is given by
\begin{align}
\label{Lpart}
 L &= [\v k \cdot  \dot{\v x} + \v A(\v x) \cdot  \dot{\v x} 
+ \v {\t A}(\v k) \cdot  \dot{\v k}] -H(\v k) -V(\v x) .
\end{align}
Here $\v A(\v x)$ is the real space vector potential for electromagnetic field
that only depends on $\v x$.  $\v{\t A}(\v k)$ is the $\v k$-space vector
potential that is assumed to depend only on $\v k$.  Such a $\v k$-space vector
potential can appear for an electron in a crystal with spin orbital couplings.
The corresponding equation of motion is given by
\begin{align}
 \dot k_i  &= -\frac{\prt V}{\prt x^i} + B_{ij} \dot x^j , &
\dot x^i &= \frac{\prt H}{\prt k_i} - \t B^{ij} \dot k_j ,
\nonumber\\
\text{or }\ \ 
\dot{\v k}  &= -\frac{\prt V}{\prt \v x} +  \dot{\v x} \times \v B , &
\dot{\v x} &= \frac{\prt H}{\prt \v k} -  \dot{\v k} \times \t{\v B} .
\end{align}
where
\begin{align}
 B_{ij} &= \prt_{x^i} A_j - \prt_{x^j} A_i , &
 \t B^{ij} &= \prt_{k_i} \t A^j - \prt_{k_j} \t A^i ,
\nonumber\\
\text{or }\ \ \v B &= \prt_{\v x}\times \v A, &
\t{\v B} &= \prt_{\v k}\times \t{\v A} .
\end{align}

Now consider a many-fermion system which is described by a particle-number
distribution $g(\xi^I)$.  The meaning of the distribution $g(\xi^I)$ is given
by
\begin{align}
 \dd N = g(\xi^I) \text{Pf}[b(\xi^I)]\frac{\dd^{2d} \xi^I}{(2\pi)^{d}} ,
\end{align}
where $\dd N$ is the number of fermions in the phase-space volume
$\dd^{2d} \xi^I$, and $\text{Pf}[b(\xi^I)]$
is the Pfaffian of the $2d\times 2d$ anti-symmetric matrix
\begin{align}
[ b(\xi^I) ]_{IJ} = b_{IJ}(\xi^I) .
\end{align}
In fact $g(\xi^I)$ have a meaning as the occupation number per orbital, since
the  number of orbitals (\ie the single particle quantum states) in the
phase-space volume $\dd^{2d} \xi^I$ is given by $
\text{Pf}[b(\xi^I)]\frac{\dd^{2d} \xi^I}{(2\pi)^{d}}$.

The above interpretation is correct since under the time evolution \eq{xiHb},
the scaled phase-space volume $\text{Pf}[b(\xi^I)]\frac{\dd^{2d}
\xi^I}{(2\pi)^{d}} $ is time independent, which corresponds to the unitary time
evolution in quantum theory.  To show such a result, we first choose a phase
space coordinate such that $b_{IJ}$ is uniform in the phase space.  In this
case, the time evolution $\dot \xi^I$ is described by a divergent-less vector
field, $\frac{\prt  H}{\prt \xi^I}$, in the phase space, and the phase-space
volume $\dd^{2d} \xi^I$ is time independent.  We note that the phase-space
volume given by the combination $\text{Pf}[b(\xi^I)]\frac{\dd^{2d}
\xi^I}{(2\pi)^{b}} $ invariant under the coordinate transformation.  Such an
invariant combination is invariant under the time evolution \eq{xiHb} for a
general coordinate, since the equation of motion is covariant under the
coordinate transformation.  We see that the phase space has a simpletic
geometry.

For our example \eq{Lpart}, the 6-by-6 matrix $b_{IJ}$ is given by
\begin{align}
 (b_{IJ}) = 
\begin{pmatrix}
B_{ij} & \del_{ij} \\
-\del_{ij}     & \t B^{ij} \\
\end{pmatrix}
\end{align} 
We find that
\begin{align}
\text{Pf}(b) &=  \text{Pf} \begin{pmatrix}
B_{ij} & \del_{ij} \\
-\del_{ij}     & \t B^{ij} \\
\end{pmatrix}
\nonumber\\
& =
\text{Pf}(B,\t B)
=1+B_{ij} \t B^{ji} + O(B_{ik}\t B^{kj})^2.
\end{align}

The effective theory for the Fermi liquid in such a general setting is simply a
hydrodynamical theory for an incompressible fluild in the phase space.  In the
following, we will write down such a theory for small fluctuations near the
ground state.  First, the ground state of the Fermi liquid is described by the
following distribution (or phase-space density)
\begin{align}
 \bar g(\xi^I) =
\begin{cases}
 1, & \text{ for } H(\xi^I) <0, \\
 0, & \text{ for } H(\xi^I) >0.\\
\end{cases}
\end{align}
The generalized Fermi surface is the $(2d-1)$-dimensional sub-manifold in the
phase space where $\bar g(\xi^I)$ has a jump.  A many-body collective
excitation is described by another incompressible distribution $g(\xi^I) =
0,1$.  For low energy collective excitations near the ground state, we may
describe such an incompressible distribution via the displacement of the
generalized Fermi surface 
\begin{align}
u(\xi^I_F) =
\sqrt{\sum_I (\Del \xi^I)^2} , 
\end{align}
where $\xi^I_F$ parametrize the $2d-1$-dimensional generalized Fermi surface,
and $\Del \xi^I$ describe the shift of the generalized Fermi surface in the
normal direction.

Let us introduce an integration over the generalized Fermi surface
\begin{align}
 \int  \text{Pf}[b(\xi^I)] \frac{\dd^{2d-1} \xi_F}{(2\pi)^d}
=: \int  \text{Pf}[b(\xi^I)]\frac{\dd^{2d} \xi}{(2\pi)^d}\,  |\prt_{\xi^I} \bar g|
\end{align}
The number of fermions in the collective excited state described by $u(\xi^I)$
is given by
\begin{align}
\label{Nuxi}
 N &= \int  \text{Pf}(b)\frac{\dd^{2d} \xi}{(2\pi)^d}\,  g(\xi^I) 
\nonumber\\
&=\bar N +
 \int  \text{Pf}(b)\frac{\dd^{2d-1} \xi_F}{(2\pi)^d}\, u(\xi^I_F)
\end{align}
The energy of the  collective excited state 
is given by
\begin{align}
\label{Euxi}
 E &=\bar E +
 \int  \text{Pf}(b)\frac{\dd^{2d-1} \xi_F}{(2\pi)^d}\, 
\frac12 |h_.|
u^2(\xi^I_F)
\end{align}
where 
\begin{align}
 h_I =: \prt_{\xi^I} H,\ \ \ \ \
|h_.| =: \sqrt{ \sum_I h_I^2}.
\end{align}

The equation of the motion of $u(\xi^I_F,t)$ can be obtained in two ways. First,
we note that $|h_I|u$ is the single particle energy, which is invariant under
the single particle time evolution $\xi^I_F(t)$ that satisfies the
single-particle equation of motion \eq{xiHb}.  Thus
\begin{align}
\frac{\dd}{\dd t}  |h_.| u(\xi^I_F(t),t) =0,
\end{align}
This allows us to obtain the equation of motion for $u(\xi^I_F,t)$ field
using the single-particle equation of motion \eq{xiHb}
\begin{align}
\label{EOMuxi1}
 \Big(\prt_t  + h^I(\xi^I_F) \prt_{\xi^I} \Big) |h_.|(\xi^I_F) u(\xi^I_F,t) =0 ,
\end{align}
where 
\begin{align}
h^I = b^{IJ}h_J
\end{align}
and the repeated index $J$ is summed.
Here $b^{IJ}$ is the matrix inversion of of $b_{IJ}$: 
\begin{align}
b_{IJ}b^{JK} = \del_{IK} .
\end{align}

Second, we note that $\text{Pf}(b) u$ is the density of fermions on the
generalized Fermi surface (see \eqn{Nuxi}).  The corresponding current density
is given by $h^I\text{Pf}(b) u$, since $\dot \xi^I = h^I$ (see \eqn{xiHb}).
The fermion conservation gives us another equation of motion for $u(\xi^I_F,t)$:
\begin{align}
\label{EOMuxi2}
 \prt_t   \text{Pf}\big(b(\xi^I_F)\big) u(\xi^I_F,t)
+\prt_{\xi^I} \Big(
h^I(\xi^I_F) \text{Pf}\big(b(\xi^I_F)\big) u(\xi^I_F,t)
\Big)
 = 0
.
\end{align}
Since the single-particle dynamics leads to
the two equations, so they they must
be consistent. This
requires that
\begin{align}
|h.| (\prt_t   +  \prt_{\xi^I} h^I) \text{Pf}(b) u
=
\text{Pf}(b) (\prt_t   +  h^I \prt_{\xi^I} ) |h.| u.
\end{align}
In other words, $h^I$, $|h.|$, and $\text{Pf}(b)$ are related, and they satisfy
\begin{align}
\label{dxih}
\frac{|h.|}{\text{Pf}(b)} 
(\prt_t   +  \prt_{\xi^I} h^I) 
\frac{\text{Pf}(b) }{|h_.|}
=
\prt_t   +  h^I \prt_{\xi^I} . 
\end{align}

Let us introduce a scalar field $\phi(\xi^I_F,t)$ via
\begin{align}
\label{uphixi}
 - |h_.|^{-1}h^I \prt_{\xi^I_F} \phi(\xi^I_F,t) = u(\xi^I_F,t),
\end{align}
The equation of motion for $\phi$ is given by
\begin{align}
\label{EOMphixi}
 (\prt_t  + h^I \prt_{\xi^I} ) h^J \prt_{\xi^I_F} \phi =0 ,
\end{align}
which can be simplified further as
\begin{align}
\label{EOMphixi}
 (\prt_t  + h^I \prt_{\xi^I} ) \phi =0 .
\end{align}
since $h^I(\xi^I_F)$ does not depend on time.

The above equation of motion and the expression of total energy
\eq{Euxi} allow us to determine the phase-space Lagrangian
\begin{align}
\label{LphFxi1}
& L_\text{ph} = 
  - \int   
\frac{\dd^{2d-1} \xi_F}{(2\pi)^d}\, 
\frac {\text{Pf}(b)} {2|h_.|}  
\Big( 
%\frac{\bar N}{VA_F}\prt_t \phi
\dot \phi h^I \prt_{\xi^I_F}  \phi 
 + [h^I \prt_{\xi^I_F}  \phi]^2 \Big) ,
\end{align}
up to total derivative topological terms.

To include topological terms, we assume a symmetry described by a map in phase
space
\begin{align}
 \xi^I \to \bar \xi^I, \ \ \ \ 
h^I(\xi^I) = -h^I(\bar \xi^I),
\end{align}
which generalize the $\v k_F \to -\v k_F$ symmetry
used before.
The phase-space Lagrangian can now be written as
\begin{align}
\label{LphFxi}
 L_\text{ph} &= 
\int   \frac{\dd^{2d-1} \xi_F}{(2\pi)^d}\, 
\Big[
\frac{\bar N}{VA_F}\dot \phi(\xi^I,t) 
\nonumber\\
&\ \ \ \ \ \ \
-
\frac {\text{Pf}\big(b(\xi^I)\big)} {2|h_.(\xi^I)|}  
\Big( 
\dot \phi(\bar \xi^I,t) h^I(\xi^I) \prt_{\xi^I_F}  \phi (\xi^I,t)
\\
+ & \dot \phi(\xi^I,t) h^I(\xi^I) \prt_{\xi^I_F}  \phi (\xi^I,t)
 + \big[h^I(\xi^I) \prt_{\xi^I_F}  \phi(\xi^I,t)\big]^2 \Big) 
\Big]
,
\nonumber 
\end{align}
where $\bar N$ is the number of fermnions in the ground state, and
\begin{align}
 VA_F = 
\int   \frac{\dd^{2d-1} \xi_F}{(2\pi)^d}\, 1
\end{align}
is the total volume of generalized Fermi surface.  

Fro the first three terms in \eqn{LphFxi}, we see that
$\dot \phi(\xi^I,t)$ directly couples to
the total density of fermions
\begin{align}
N = 
\int   \frac{\dd^{2d-1} \xi_F}{(2\pi)^d}\, 
\Big[
\frac{\bar N}{VA_F} &
-
\frac {\text{Pf}\big(b(\xi^I)\big)}{2|h_.(\xi^I)|}  
\Big( 
  h^I(\xi^I) \prt_{\xi^I_F}  \phi (\xi^I,t)
\nonumber \\
& - h^I(\xi^I) \prt_{\bar \xi^I_F}  \phi (\bar \xi^I,t)
\Big) 
\Big]
.
\end{align}
In particular, the uniform part of $\dot \phi(\xi^I,t)$
couple to total number of fermions
\begin{align}
 L_\text{ph} = N \dot \phi_\text{uniform}(t) + \cdots .
\end{align}
This indicates that $\phi \sim \phi +2\pi$ is an angular field, and the $U(1)$
transformation is given by
\begin{align}
\phi(\xi^I,t) \to \phi(\xi^I,t)+\th . 
\end{align}

\subsection{Effective theory for a Fermi liquid with real space and $\v
k$-space magnetic fields}

Now, let us apply the above formalism to develop the low energy effective
theory of Fermi liquid, for 3-dimensional fermions with real space magnetic
field $\v A(\v x)$ and $\v k$-space ``magnetic'' field $\t{\v A}(\v k)$.  The
dynamics of a single fermion is described by \eqn{Lpart} (with $V=0$).  
%Up to
%second order in $\v B $ and $\t{\v B} $, 
We have
\begin{align}
 (h_I) &= (0, {\v v}_F), \ \ \ \ \ |h_.| = |\v v_F|, \ \ 
{\v v}_F =: \prt_{\v k} H,
\nonumber\\
 (h^I) &= (\t{\v v}_F , \v f), \ \ \ \ \
\text{Pf}(b)   = 1 -2 \v B\cdot \t{\v B} +\cdots,
\\
 \t{\v v}_F &= {\v v}_F - ({\v v}_F\times \v B)\times \t{\v B} + \cdots, 
\ \
 \v f =: {\v v}_F\times \v B +\cdots .
\nonumber 
\end{align}
Here $\v f = \dot{\v k}$ has a physical meaning as the force acting on each
fermion.  $\t{\v v}_F = \dot{\v x}$ has a physical meaning as the velocity of
each fermion.  When $\t{\v A} \neq 0$, the velocity of a fermion at the Fermi
surface is not given by $\v v_F=\prt_{\v k} H$.  $\t{\v v}_F$ is also called
the anomalous velocity.\cite{SNc9908003}

Substitute the above into \eqn{LphFxi}, we obtain
\begin{widetext}
\begin{align}
\label{LphFBtB}
L_\text{ph} = 
 \int \dd^d \v x & \frac{\dd^{d-1} \v k_F}{(2\pi)^d}\, 
\Big(  \frac{\bar \rho}{A_F}  \dot \phi (\v x,\v k_F)
-\frac{\text{Pf}(b) \dot \phi (\v x,-\v k_F)}{2|\v v_F|}
( \t{\v v}_F \cdot \prt_{\v x} +  \v f \cdot \prt_{\v k_F} ) 
\phi (\v x,\v k_F,t)
\\
& -\frac{\text{Pf}(b)\dot \phi (\v x,\v k_F)}{2|\v v_F|}
( \t{\v v}_F \cdot \prt_{\v x} +  \v f \cdot \prt_{\v k_F} ) 
\phi (\v x,\v k_F,t)
 -
\frac{\text{Pf}(b)}{2|\v v_F|}
\big[
( \t{\v v}_F \cdot \prt_{\v x} +  \v f \cdot \prt_{\v k_F} ) 
\phi (\v x,\v k_F,t)
\big]^2
\Big)
\nonumber\\
& \hskip -0.6in  - 
\int \dd^d \v x  \frac{\dd^{d-1} \v k_F}{(2\pi)^d}
 \frac{\dd^{d-1} \v k_F'}{(2\pi)^d} \,
 \, 
\frac{\text{Pf}(b)\text{Pf}(b')V(\v k_F,\v k_F')}{2|\v v_F| |\v v_F'|}
\Big(
( \t{\v v}_F \cdot \prt_{\v x} +  \v f \cdot \prt_{\v k_F}) 
\phi (\v x,\v k_F,t) 
\Big)
\Big(
( \t{\v v}_F' \cdot \prt_{\v x} +  \v f^{\prime} \cdot \prt_{\v k_F} )
\phi (\v x,\v k_F',t) 
\Big)
.
\nonumber 
\end{align}
Up to first order in $\v B $ and
$\t{\v B}$, the above can be simplified
\begin{align}
\label{LphFB}
L_\text{ph} = 
 \int \dd^d \v x & \frac{\dd^{d-1} \v k_F}{(2\pi)^d}\, 
\Big(  \frac{\bar \rho}{A_F}  \dot \phi (\v x,\v k_F)
-\frac{ \dot \phi (\v x,-\v k_F)}{2|\v v_F|}
( \v v_F \cdot \prt_{\v x} +  \v f \cdot \prt_{\v k_F} ) 
\phi (\v x,\v k_F,t)
\\
& -\frac{\dot \phi (\v x,\v k_F)}{2|\v v_F|}
( \v v_F \cdot \prt_{\v x} +  \v f \cdot \prt_{\v k_F} ) 
\phi (\v x,\v k_F,t)
 -
\frac{1}{2|\v v_F|}
\big[
( \v v_F \cdot \prt_{\v x} +  \v f \cdot \prt_{\v k_F} ) 
\phi (\v x,\v k_F,t)
\big]^2
\Big)
\nonumber\\
& \hskip -0.6in  - 
\int \dd^d \v x  \frac{\dd^{d-1} \v k_F}{(2\pi)^d}
 \frac{\dd^{d-1} \v k_F'}{(2\pi)^d} \,
 \, 
\frac{V(\v k_F,\v k_F')}{2|\v v_F| |\v v_F'|}
\Big(
( \v v_F \cdot \prt_{\v x} +  \v f \cdot \prt_{\v k_F}) 
\phi (\v x,\v k_F,t) 
\Big)
\Big(
( \v v_F' \cdot \prt_{\v x} +  \v f^{\prime} \cdot \prt_{\v k_F} )
\phi (\v x,\v k_F',t) 
\Big)
.
\nonumber 
\end{align}
\end{widetext}
Here, we have assumed a central reflection symmetry $\v k \to -\v k$, and the
mapping $\xi^I \to \bar \xi^I$ is given by $(\v x, \v k) \to (\v x, -\v k)$.
We also included the interaction term for the Fermi surface fluctuations, the
$V(\v k_F,\v k_F')$ term, where $\v v_F, \v v_F'$ are the Fermi velocities at
$\v k_F, \v k_F'$.

Note that the fermion density at the Fermi surface $\v k_F$ is given by (see
\eqn{uphixi})
\begin{align}
 u(\v x,\v k_F) = -
\frac{ \v v_F}{|\v v_F|} \cdot \prt_{\v x} \phi (\v x,\v k_F,t) 
-  \frac{\v f}{|\v v_F|} \cdot \prt_{\v k_F} \phi (\v x,\v k_F,t),
\end{align}
and thus the total fermion number density is given by
\begin{align}
 \rho  = \bar \rho -
\int \frac{\dd^{d-1} \v k_F}{(2\pi)^d}\, 
\Big( 
\frac{ \v v_F}{|\v v_F|} \cdot \prt_{\v x} \phi  
+  \frac{\v f}{|\v v_F|} \cdot \prt_{\v k_F} \phi
\Big)
.
\end{align}
This expression helps us to understand why the interaction term for the Fermi
surface fluctuations has a form given in \eqn{LphFB}.

The above expression also allows us to see that $\dot \phi(\v x,\v k_F)$
couples to the total fermion density (see the first three terms in \eqn{LphFB}).
Thus $\phi(\v x,\v k_F) \sim \phi(\v x,\v k_F) +2\pi$ is an angular field
and the $U(1)$ symmetry transformation is given by
\begin{align}
 \phi(\v x,\v k_F) \to \phi(\v x,\v k_F) +\th .
\end{align}

Eqn. \eq{dxih} now becomes (to the first order in
$\v B$ and $\t{\v B}$)
\begin{align}
\label{vvphi}
\prt_{\v x} \cdot \v v_F +  \prt_{\v k_F} \cdot \v f 
=
|\v v_F|^{-1} 
\big( 
\v v_F \cdot \prt_{\v x} +  \v f \cdot \prt_{\v k_F} 
\big) 
|\v v_F| 
\end{align}
This will help us to compute the equation of motion for the $\phi$ field.  The
resulting equation of motion is given by (written in terms of $u(\v x,\v k_F,t)$)
\begin{widetext}
\begin{align}
\label{BEA}
\big(\prt_t + \v v_F \cdot \prt_{\v x}  +  \v f \cdot \prt_{\v k_F} \big) 
|\v v_F| u(\v x,\v k_F,t) 
 +   \big(\v v_F \cdot \prt_{\v x}  +  \v f  \cdot \prt_{\v k_F} \big)  
\int \frac{\dd^{d-1} \v k_F'}{(2\pi)^d} \, 
V(\v k_F,\v k_F') u(\v x, \v k_F',t)
=0.
\end{align}
The above is the Boltzmann equation which can be use to compute the transport
properties after adding the collision terms.  In terms of $\phi(\v x,\v
k_F,t)$, we have
\begin{align}
\label{EOMphiA}
\big(\prt_t + \v v_F \cdot \prt_{\v x}  +  \v f \cdot \prt_{\v k_F} \big)  \phi(\v x,\v k_F,t) 
+ \int \frac{\dd^{d-1} \v k_F'}{(2\pi)^d}
\, 
\frac{V(\v k_F,\v k_F')}{|\v v_F'|}
\big( \v v_F' \cdot \prt_{\v x} \phi (\v x,\v k_F',t) 
+  \v f^{\prime} \cdot \prt_{\v k_F} \phi (\v x,\v k_F',t) \big)
=0 .
\end{align}
The above equations of motion are valid only to the first order in $\v B$ and
$\t{\v B}$.  The exact equations of motion, in several different forms, are
given by
\begin{align}
\label{BEAtA}
&\big(\prt_t + \prt_{\v x}  \cdot \t{\v v}_F +  \prt_{\v k_F} \cdot \v f \big) 
\text{Pf}(b) u(\v x,\v k_F,t) 
 + \text{Pf}(b) 
\big(\frac{\t{\v v}_F}{|\v v_F|} \cdot \prt_{\v x}  
+  \frac{\v f}{|\v v_F|}  \cdot \prt_{\v k_F} \big)  
\int \frac{\dd^{d-1} \v k_F'}{(2\pi)^d}
\, 
\text{Pf}(b')
V(\v k_F,\v k_F')
u(\v x, \v k_F',t)
=0,
\nonumber \\
&\big(\prt_t + \t{\v v}_F \cdot \prt_{\v x}  +  \v f \cdot \prt_{\v k_F} \big) 
|\v v_F| u(\v x,\v k_F,t) 
 +    \big(\t{\v v}_F \cdot \prt_{\v x}  
 +  \v f  \cdot
\prt_{\v k_F} \big)  \int \frac{\dd^{d-1} \v k_F'}{(2\pi)^d}
\, 
\text{Pf}(b')
V(\v k_F,\v k_F')
u(\v x, \v k_F',t)
=0,
\\
&\big(\prt_t + \t{\v v}_F \cdot \prt_{\v x}  +  \v f \cdot \prt_{\v k_F} \big)  \phi(\v x,\v k_F,t) 
+ \int \frac{\dd^{d-1} \v k_F'}{(2\pi)^d} \, 
\frac{\text{Pf}(b')}{|\v v_F'|}V(\v k_F,\v k_F')
\big( \v v_F' \cdot \prt_{\v x} \phi (\v x,\v k_F',t) 
+  \v f^{\prime} \cdot \prt_{\v k_F} \phi (\v x,\v k_F',t) \big)
=0 .
\nonumber 
\end{align}
\end{widetext}

\subsection{Emergent $U^\infty(1)$ symmetry}

From the effective theory \eq{LphFBtB}, we see when there is no real space
magnetic field $\v B=0$, we have $\v f=0$ and the  effective theory has a
$U^\infty(1)$ symmetry generated by
\begin{align}
 \phi(\v x, \v k_F) \to \phi(\v x, \v k_F) + \th(\v k_F) ,
\end{align}
where $\th(\v k_F)$ can be any function of $\v k_F$.  This is the so called
emergent $U^\infty(1)$ symmetry, which a key character of Fermi
liquid.\cite{L7920,Hc0505529,HM9390,NF9493,ES200707896} We also see that the
above transformation is no longer a symmetry in the presence of real space
magnetic field $\v B\neq 0$.  This may be related to the anomaly in the
emergent $U^\infty(1)$ symmetry discussed in \Ref{ES200707896}

\section{Fermion-pair liquid and the mixed $U(1)\times \R^d$ anomaly}
\label{FpL}

In this section, we are going to consider a fermion system in $d$-dimensional
continuous space, with $U(1)$ particle-number-conservation symmetry and $\R^d$
translation symmetry.  We assume the space to have a size $L_1\times L_2\times
\cdots \times L_d$, and has a periodic boundary condition.  We will compute
distribution of the total momentum for many-body low energy excitations $\v
k_\text{tot}$, and how such a distribution depends on the $U(1)$ symmetry twist
described by a constant vector potential $\v a$.

Using the results from the Appendix \ref{Var}, we find that the low energy
effective theory is described by the following phase-space Lagrangian (see
\eqn{Lx})
\begin{align}
\label{LFp}
  L = \int \dd^d \v x\;  &
\big(
 \bar \rho_p \dot \phi_p(\v x,t) +
\del \rho_p (\v x,t) \dot \phi_p(\v x,t)
\nonumber\\
&\ 
 - \frac {\bar \rho}{2M_p} |\v \prt \phi_p|^2 
-  \frac{g}{2} \del \rho_p^2 +\ \cdots 
\big) ,
\end{align}
where $\bar \rho_p = \frac12 \bar \rho$ is the fermion-pair density in the
ground state, and $\phi_p \sim \phi_p +2\pi$ is the angular field for the
fermion-pair.  The total energy and total crystal momentum of those excitations
are given by
\begin{align}
\label{EKLmu}
 E &= \frac{\bar N_p}{2M_p} \sum_\mu (\frac{m_\mu}{L_\mu} + 2 a_\mu )^2 +
\sum_{\v k\neq 0} (n_{\v k} +\frac12) v |\v k|,
\nonumber\\
\v k_\text{tot} &= N_p \sum_\mu (\frac{m_\mu}{L_\mu} + 2 a_\mu )\hat{\v x}_\mu +
\sum_{\v k\neq 0} n_{\v k} \v k,
\end{align}
We see that the $U(1)$ symmetry twist $\v a$ induces a change in the total
momentum 
\begin{align}
\label{krhoP}
\v k_\text{tot}
= 2 N_p \v a 
=  2 \bar \rho_p \v a V
=  \bar \rho \v a V.
\end{align}
Such momentum dependence of the $U(1)$ symmetry twist $\v a$ reflects the mixed
$U(1)\times \R^d$ anomaly.  Eqn. \eq{krhoP} and \eqn{krho} are identical,
implies the identical mixed anomaly, which is captured by the topological term
$\int \dd^d \v x\, \big( \bar \rho_p \dot \phi_p(\v x,t) $.  

The mixed anomaly can also be measured by the periodicy $k_{0\mu}$ in the
distribution of $\v k_\text{tot}$, and the periodicy $\Del N=2$ in the
distribution of $N$, for  the low energy excitations.  The period in
$\mu$-direction times $\Del N$ is
\begin{align}
\Del N
k_{0\mu} = 2\frac{N_p}{L_\mu} = \frac{N}{L_\mu}.
\end{align}
The periodices $k_{0\mu}$ and $\Del N$ are universal low energy properties of
the fermion-pair gapless phase.  Their product $\Del N k_{0\mu}$ is even more
robust, since it is invariant even across any phase transitions, and thus
correspond to an anomaly.

For the gapless state formed by four-fermion bound states, the periodicies will
be
\begin{align}
k_{0\mu} = \frac14  \frac{N}{L_\mu},\ \ \ \ \ \ \ \ \Del N= 4.
\end{align}
Their product is still $\Del N k_{0\mu}  = \frac{N}{L_\mu}$.

~

~

I would like to thank Maissam Barkeshli, Dominic Else, and Senthil Todadri for
discussions and comments.  This research is partially supported by NSF
DMR-2022428 and by the Simons Collaboration on Ultra-Quantum Matter, which is a
grant from the Simons Foundation (651440).

\appendix

\section{Dynamical variational approach and low energy effective theory}
\label{Var}

\subsection{Coherent state approach}

A quantum state is described by a complex vector
\begin{equation*}
|\psi\> = \bpm
\psi_1\\
\psi_2\\
\vdots\\
\epm
\end{equation*}
in a Hilbert state,
with inner product
\begin{align}
 \<\phi|\psi\> = \sum_n \phi^*_n \psi_n.
\end{align}
The motion of a quantum state is described by time
dependent vector: $|\psi(t)\>$, which satisfy an equation of motion (called
Shr\"odinger equation) with only first order time derivative:
\begin{align}
\label{Seqn}
\ii \frac{\dd}{\dd t} |\psi(t)\> = \hat H |\psi(t)\> ,
\end{align}
where the hermitian operator $\hat H$ is the Hamiltonian.

The Shr\"odinger equation \eq{Seqn} also has a phase-space Lagrangian
description. If we choose the Lagrangian $L$ to be
\begin{align}
\label{QLag}
 L(\frac{\dd}{\dd t}|\psi\>, |\psi\>) &=\<\psi(t)|\ii \frac{\dd}{\dd t}
  -\hat H |\psi(t)\>
\nonumber\\
&= \ii \psi_n^*(t) \dot \psi_n(t)
- \psi_m^*(t) H_{mn} \psi_n(t),
\end{align}
then the action $S=\int \dd t\, L(\frac{\dd}{\dd t}|\psi\>, |\psi\>)$ will be a
functional for the paths $|\psi(t)\>$ in the Hilbert space.  The stationary
paths $|\psi_\text{sta}(t)\>$ of the action will correspond to the solutions of
the Shr\"odinger equation.  Since the Shr\"odinger equation can be derived from
the Lagrangian, we can say that the Lagrangian $L(\frac{\dd}{\dd t}|\psi\>,
|\psi\>)$ provides a complete description of a quantum system.

In the variational approach to the ground state, we consider a variational
state $|\psi_{\xi^i}\>$ that depends on variational parameters $\xi^i$.  We
then found an approximation of the ground state $|\psi_{\bar \xi^i}\>$ by
choosing $\bar \xi^i$ that minimize the average energy
\begin{align}
\bar H(\xi^i)=  \< \psi_{\xi^i}| \hat H |\psi_{\xi^i}\>.
\end{align}

If we choose the variational parameters $\xi^i$ properly, the low energy
excitations are also described by the fluctuations of variational parameters.
In other words, the dynamics of the variational parameters $\xi^i$ described
the  low energy excitations. This leads to a dynamical variational approach (or
coherent state approach) that gives us a description of both ground state and
low energy excitations.

The  dynamics of the full quantum system  is described by phase-space
Lagrangian $L=\<\psi|\ii \frac{\dd}{\dd t} -\hat H|\psi\>$.  The  dynamics of
the variational parameters is described by the evolution of the quantum states
in a submanifold of the total Hilbert space, given by the variational states $
|\psi_{\xi^i}\>$.  Here we want to obtain the dynamics of the quantum states,
restricted to the submanifold parametrized by $\xi^i$.  Such a dynamics is
described by the same phase-space Lagrangian restricted in the submanifold:
\begin{equation}
\label{Lxi}
 L(\dot \xi^i,\xi^i)=
\<\psi_{\xi^i(t)}|
\ii \frac{\dd}{\dd t} -\hat H |\psi_{\xi^i(t)}\>
=a_i(\xi^i) \dot \xi^i - \bar H(\xi^i)
\end{equation}
where
\begin{equation}
\label{axi}
 a_i=\ii \<\psi_{\xi^i}| \frac{\prt}{\prt \xi^i}|\psi_{\xi^i}\>
,\ \ \ \ \ \ 
\bar H(\xi^i)=\<\psi_{\xi^i}| \hat H |\psi_{\xi^i}\>
\end{equation}
The resulting equation of motion is given by
\begin{equation}
\label{genHeompsi}
 b_{ij}\dot \xi^j = \frac{\prt \bar H}{\prt \xi^i}, \ \ \ \ \ \ \
 b_{ij}=\prt_i a_j -\prt_j a_i
\end{equation}
which describes the classical motion of $\xi^i$.

The above phase-space Lagrangian actually only described the classical dynamics
of the variables $\xi^i$.  The obtain the low energy effective theory for the
quantum dynamics of the  variables $\xi^i$, we need quantize the  phase-space
Lagrangian \eq{Lxi} to obtain the low energy effective Hilbart space
$\cH_\text{eff}$ and the low energy effective Hamiltonian $H_\text{eff}$ acting
with $\cH_\text{eff}$.
Roughly,
the low energy effective Hilbart space
$\cH_\text{eff}$
is an representation
of the operators algebra
\begin{align}
\ii [\hat \xi^i, \hat \xi^j ] = b^{ij}(\hat \xi^i),
\end{align}
where $b^{ij}$ is the inverse of $b_{ij}$: $b_{ij} b^{jk} =\del_{ik}$.  The low
energy effective Hamiltonian is given by 
\begin{align}
H_\text{eff} = \bar H(\hat \xi^i).
\end{align}

Let us use the above approach to describe hardcore boson on a single site.
The total Hilbert space is 2-dimensional, spanned by $|0\>$ (no boson) and
$|1\>$ (one boson).  The coherent state is described by a unit vector $\v n =
(n_x,n_y,n_z)  \in S^2$.
We may also use $\th,\phi$ to describe $\v n$:
\begin{align}
 n_x(\th,\phi) &= \cos \phi \sin \th ,
\nonumber\\
 n_x(\th,\phi) &= \sin \phi \sin \th ,
\nonumber\\
 n_z(\th,\phi) &= \cos \th .
\end{align}
We can choose the coherent state to be
\begin{align}
 |\v n(\th,\phi)\> =
\begin{pmatrix}
 \cos \frac{\th}{2}\\
 \ee^{-\ii \phi} \sin \frac{\th}{2}\\
\end{pmatrix}
\end{align}
The phase-space Lagrangian to describe the classical dynamics
of $\th,\phi$ is given by
\begin{align}
L & = \sin^2 \frac{\th}{2} \dot \phi - \bar H(\th,\phi)
\nonumber\\
&= 
\rho \dot \phi - \bar H(\rho,\phi),\ \ \ \ \ 
\rho  \equiv \frac{1-n_z}{2} =\sin^2 \frac{\th}{2}
\end{align}
We will use $(\rho,\phi)$ to parametrize the phase space, where $\rho \in
[0,1]$ has a physical meaning being the average number of bosons on the site.
Here we stress that $(\rho,\phi)$ parametrize $S^2$.  Quantizing the above
classical phase-space Lagrangian, we suppose to obtain a quantum system with
Hilbert space $\cH = \text{span}(|0\>,|1\>)$.

\subsection{Low energy effective theory of bosonic superfluid phase}

Using the above result, we obtain the following low energy effective theory for
interacting bosons in a $d$-dimensional cubic lattice with periodic boundary
condition, whose sites are labeled
by $\v i$:
\begin{align}
 L = \sum_{\v i} \rho_{\v i} \dot \phi_{\v i} 
- \bar H(\rho_{\v i},\phi_{\v i}),
\end{align}
where $\phi_{\v i} \sim \phi_{\v i}+2\pi$ is an angular variable.
If we assume $ \rho_{\v i}, \phi_{\v i}$ to have a smooth dependence 
on the space coordinate $\v x \sim \v i$, the above can be rewritten as a
field theory
\begin{align}
\label{Lx}
  L = \int \dd^d \v x\;  &
\big(
\bar \rho \dot \phi(\v x,t) +
\del \rho(\v x,t) \dot \phi(\v x,t)
\nonumber\\
&\ 
 - \frac {\bar \rho}{2m} |\v \prt \phi|^2 -  \frac{g}{2} \del \rho^2 +\ \cdots 
\big) ,
\end{align}
where we have assumed that $\bar H(\rho,\phi)$ is minimized as $\rho =\bar
\rho$ which correspond to average number of bosons per site.  
%If we only interested in low energy excitations, we can ignore the interaction term $g$.

To quantize the above low-energy-effective theory, we expand
\begin{align}
 \del \rho &= 
\rho_0 +\sum_{\v k\neq 0} \rho_{\v k} \frac{\ee^{\ii \v k \cdot \v x}}{L^{d/2}}
\nonumber\\
  \phi &= 
\phi_0 
+ 2\pi \frac{\v m \cdot \v x}{L}
+\sum_{\v k\neq 0} \phi_{\v k} \frac{\ee^{\ii \v k \cdot \v x}}{L^{d/2}},
\end{align}
where $L$ is the size of the cubic lattice.
The $\v k\neq 0$ modes give rise to a collection of quantum oscillators after
quantization.  $\rho_0,\phi_0, \v m$ describe the $\v k=0$ mode, where the
integer vector $\v m=(m_1,\cdots,m_d)$ describes the winding numbers of the
phase $\phi$.  The effective Lagrangian for the $\v k=0$ modes is given by
\begin{align}
  L_0 = 
\big( L^d
\bar \rho \dot \phi_0 + L^d
\rho_0 \dot \phi_0
 - \frac {\bar \rho L^{d-2}}{2m}  (2\pi)^2 \v m^2 - L^d \frac{g}{2} \rho_0^2
\big) .
\end{align}
After quantization, the  $\v k=0$ mode describes a particle on a ring with $L^d
\bar \rho$ flux through the ring.
Let $\hat L_z \sim L^d (\bar\rho +\rho_0)$ be the angular operator of the
quantized particle.
After quantization, the Hamiltonian
is given by
\begin{align}
 \hat H_0 = \frac {\bar \rho L^{d-2}}{2m}  (2\pi)^2\v m^2 
+ \frac{g}{2 L^d} (\hat L_z - L^d \bar \rho)^2
\end{align}

The many-body low energy excitations are labeled by $(\v m, N, n_{\v k\neq 0})$,
where integer  $N$ is the eigenvalues of $\hat L_z$ (the total number of
bosons) and integer $n_{\v k\neq 0})$ is the number of excited phonons for $\v
k$ mode.  The total energy and the total crystal momentum are given by
\begin{align}
&E = \bar N \frac {(2\pi)^2 \v m^2/L^2}{2m }   
+ \frac g2 \frac{(N -\bar N)^2}{ L^d} 
+ \sum_{\v k\neq 0} (n_{\v k}+\frac12) v |\v k|^2,
\nonumber\\
&\v k_\text{tot} = \int \dd^d \v x\; (\bar \rho+\del \rho)  \v \prt \phi
=N \frac {2\pi \v m}{L} +  \sum_{\v k\neq 0} n_{\v k} \v k
\end{align}
where the phonon velocity $v = \sqrt{\frac{g \bar \rho}{m}}$, and
$\bar N \equiv L^d \bar \rho_0$.

The above results are very standard, except that we carefully keep the
topological term $\bar \rho(\v x,t) \dot \phi(\v x,t)$ in the Lagrangian.  Our
quantum system has $U(1)$ particle number conservation symmetry and $\Z^d$
lattice translation symmetry.  In the continuum field theory \eq{Lx}, we take
the limit of zero lattice spacing.  In this case, both the $U(1)$ and $\Z^d$
symmetries are internal symmetries of the field theory.  It turns out that the
$U(1)\times \Z^d$ symmetry has a mixed 't Hooft anomaly when $\bar \rho$ is not
an integer, which constrains the low energy dynamics of the interacting bosons.

\bibliography{../../bib/all,../../bib/allnew,../../bib/publst}
\end{document}